# Nonlinear Dynamics of Spherical Shells Buckling under Step Pressure

Jan Sieber [a], John W. Hutchinson [b] and J. Michael T. Thompson [c]

[a] CEMPS, University of Exeter, Exeter EX4 4QF, UK
[b] SEAS, Harvard University, Cambridge MA 02138, USA
[c] DAMPT, University of Cambridge, Cambridge CB3 0WA, UK

**Abstract**   Dynamic buckling is addressed for complete elastic spherical shells subject to a rapidly applied step in external pressure.  Insights from the perspective of nonlinear dynamics reveal essential mathematical features of the buckling phenomena.  To capture the strong buckling imperfection-sensitivity, initial geometric imperfections in the form of an axisymmetric dimple at each pole are introduced.  Dynamic buckling under the step pressure is related to the quasi-static buckling pressure.  Both loadings produce catastrophic collapse of the shell for conditions in which the pressure is prescribed.  Damping plays an important role in dynamic buckling because of the time-dependent nonlinear interaction among modes, particularly the interaction between the spherically symmetric 'breathing' mode and the buckling mode.  In general, there is not a unique step pressure threshold separating responses associated with buckling from those that do not buckle.  Instead there exists a cascade of buckling thresholds, dependent on the damping and level of imperfection, separating pressures for which buckling occurs from those for which it does not occur.  For shells with small and moderately small imperfections the dynamic step buckling pressure can be substantially below the quasi-static buckling pressure.

## 1. Introduction

Together with the axially compressed cylindrical shell, the complete elastic spherical shell under spatially uniform external pressure is one of the two archetypal examples of shell buckling. For this reason, it is studied extensively as a proving ground for advanced shell-buckling theories, involving the notorious imperfection-sensitivity. Recently there has been a renewed interest in this problem, stimulated by recent developments in shell buckling on several fronts.  These include efforts underway in China, Europe and the U.S. to update design criteria for shell buckling accounting for advances in computation and shell manufacturing since the criteria were established in the mid-1960's. Also there have been recent advances in experimental and theoretical aspects including new experiments on spherical shells with precisely manufactured imperfections [1], accurate formulations and simulations [1,2], and proposals to assess buckling and imperfection-sensitivity by experimental probing techniques [3] executed in [4,5].

Work on spherical shell buckling so far has concentrated on the static behaviour under the slow increase of the spatially uniform loading, under both (dead) pressure control and (rigid) volume control. Particular attention has been given to the imperfection sensitivity and associated energy barriers against collapse under operating conditions. Our aim in this paper is to initiate some high-precision dynamical studies of imperfect spherical shells, following a common historical pattern by first addressing the problem of step-loading. In this, we examine the highly nonlinear dynamical response of imperfect shells when, at rest under zero load, the shell is subjected to a rapidly applied dead pressure of magnitude $p$ which then remains constant over the time, $t$, which is effectively allowed to tend to infinity. This loading process is then repeated at a fine set of different $p$ values, and a record is kept as to which values of $p$ result in a buckling collapse inferred by the dynamical response undergoing a dramatic increase in magnitude. The study limits imperfections and deformations to be axisymmetric which nevertheless captures all the essential nonlinearity of spherical shells buckling under uniform pressure. In this sense, the spherical shell is ideal for an in-depth investigation of dynamic buckling of imperfection-sensitive structures. At very large deflections, non-axisymmetric departures from the axisymmetric response can occur, though in a range far beyond that relevant to the onset of buckling [6].

In the simplest scenario at a fixed imperfection magnitude, $\delta$, the set of increasing step-loads will exhibit no collapse until some $p = p_D$ but guaranteed collapse at $p > p_D$. We can then identify $p_D(\delta)$ as the dynamic buckling load at this $\delta$, and its graph can be compared with the static imperfection sensitivity curve $p_S(\delta)$. This simple scenario does however often break down due to the dynamical phasing as the energy barrier is approached. Unlike predictions based on simpler one- and two-degree of freedom nonlinear structural systems [7], collapse is not guaranteed for $p > p_D$, but instead we observe a sequence of thresholds. Here, recent work on nonlinear dynamics provides insights into why there is not a unique threshold characterizing dynamic buckling of the shells under step pressure. We believe the concepts revealed for the spherical shell will carry over to step loading of other shell structures. Earlier work on dynamic buckling, some of which is reviewed in [7], has not revealed the complexity of dynamic buckling nor the insights offered by perspectives of nonlinear dynamics.

In the rest of this Introduction we list the content of the sections and anticipate some of the findings. The equations governing the nonlinear behaviour of complete thin, elastic spherical shells are presented in Section 2 along with a brief outline of the numerical solution methods. A summary of relevant quasi-static results for the elastic buckling of spherical shells is presented in Section 3, including the effect of imperfections on reducing the buckling pressure and the energy barrier to buckling for shells subject to sub-failure pressures. The paper focuses on prescribed spatially uniform (dead) pressure loadings where the device applying the pressure is not influenced by the deformation of the shell. Under quasi-statically increased load, buckling takes place at the maximum pressure the shell can support and is followed by complete dynamic collapse. Dimple-shaped geometric imperfections will be considered which are both realistic and among the most deleterious to buckling. The shape and amplitude of the imperfections will be scaled such that our results for buckling will be essentially independent of the all-important shell radius to thickness ratio, $R/h$, for shells with $R/h$ greater than about 25 to 50. Under step loading, the dynamic coupling between spherically symmetric breathing (vibration) mode and the incipient buckling mode plays a critical role, and thus, in Section 3, we also present selected results on these modes based on linearization about both stable unbuckled states and unstable buckled states.

Dynamic responses for step loaded shells are shown in Section 4. Comparison between the dynamic and quasi-static buckling pressures is made dependent on the imperfection amplitude. For nearly perfect shells or those with relatively modest levels of imperfection, the lowest dynamic buckling pressure falls significantly below the quasi-static buckling pressure but, conversely, falls far above lower bounds based on energy barrier concepts. For larger imperfections, the step buckling load is only slightly below the quasi-static buckling pressure. We observe a significant delay between application of pressure step and occurrence of buckling.

Especially at or just above the lowest threshold for step buckling, buckling occurs only after multiple oscillations of the breathing mode with a slow transfer of energy from the breathing mode to the buckling mode. For larger imperfections ($\delta/h > 0.25$) imperfection sensitivity trends for the step buckling pressure are similar to those for quasi-static buckling (at slightly lower pressures) and not strongly dependent on damping.

Because of the time-delay before buckling occurs, damping comes into play in the case of small imperfections and thus in the determination of the step buckling threshold. Damping in this paper arises from, and can be controlled by, the time discretization employed in the numerical simulations. Section 5 is devoted to a discussion of the role of damping on the determination of the step buckling pressures. While, for larger imperfections we will see imperfection sensitivity trends for the step buckling pressure which are similar to (at slightly lower pressures) those for quasi-static buckling and not strongly dependent on damping.

Section 6 examines the cascade of step-buckling thresholds from the perspective of work in nonlinear dynamics. Each threshold pressure corresponds to a response where the pole of the shell performs one more large oscillation before crossing the buckling threshold. This implies that the surface forming the buckling threshold in phase space has a complicated geometry folding around the unbuckled state many times.

## 2. Notation, governing equations and dimensionless quantities

This paper considers thin spherical shells of radius $R$ and thickness $h$. The shell material is isotropic and linearly elastic with Young's modulus $E$, Poisson's ratio $\nu$, and uniform density $\rho$. Geometric imperfections, $w_I$, in the location of the shell middle surface will be introduced. All numerical results in this paper are based on the small strain-moderate rotation theory (Sanders [8], Koiter [9]) for axisymmetric deformations of the shell. Specifically, we assume that motion is rotationally symmetric around the North-South pole axis with reflection symmetric about the equator. This theory, reviewed and employed for the symmetric case in [2], is accurate for thin shells, e.g., $R/h \geq 50$, if the largest deflections which occur at the poles do not exceed about $0.2R$, which will always be true in the range of interest in this paper. This paper is only concerned with whether the shell buckles or not—no attempt will be made to attain the collapse state. The onset of buckling can be determined from simulations of deflections within finite multiples of the thickness ($4$ to $6h$).

*(a) Governing equations according to small strain-moderate rotation theory*

All displacements can be described as functions of the meridional angle $\theta \in [0, \pi/2]$ ($\theta = 0$ at the equator, $\theta = \pi/2$ at the North pole, as shown in an inset in Figure 4). Rotational symmetry implies that each middle surface point's displacement on the shell has a single tangential component $u(\theta,t)$ and an outward normal component $w(\theta,t)$ in the radial direction. The middle surface strains $(\varepsilon_\theta, \varepsilon_\omega)$ and bending strains $(K_\theta, K_\omega)$ in small strain/moderate rotation theory for these conditions of symmetry are, with $\varphi = -W' + U$ as the linearized rotation,

$$\varepsilon_\theta = U' + W + \frac{1}{2}\varphi^2 - \varphi W_I', \quad \varepsilon_\omega = -U\tan\theta + W \tag{2.1}$$

$$\kappa_\theta = \varphi', \quad \kappa_\omega = -\varphi \tan\theta \tag{2.2}$$

Here, $(\ )' = \partial(\ )/\partial\theta$, and the dimensionless displacements and bending strains are $(W, U) = (w, u)/R$ and $(\kappa_\theta, \kappa_\omega) = R(K_\theta, K_\omega)$. Subscript $\theta$ refers to the meridional direction, while $\omega$ refers to the circumferential direction. An initial imperfection in the form of a normal stress-free displacement of the shell middle surface, $w_I(\theta)$, from the perfect spherical shape has dimensionless form $W_I = w_I/R$. The imperfection used in this paper is a set of identical inward dimples at each pole with shape specified by (at the upper pole with $\beta = \pi/2 - \theta$)

$$w_I = -\delta e^{-(\beta/\beta_I)^2} \quad \text{with} \quad \beta_I = B/\sqrt{\sqrt{1-\nu^2}R/h} \tag{2.3}$$

and $\delta$ as the imperfection amplitude. The radius $\beta_I$ is a measure of the width of the Gaussian shaped dimple. These imperfections are realistic with $\delta/h$ usually not larger than about unity [1,2]. The scaling of $\beta_I$ ensures that the relation between the buckling pressure and the amplitude $\delta$ is essentially independent of $R/h$ for thin shells. For a given imperfection amplitude $\delta$, there is a value of $B$ that produces the minimum buckling pressure. The choice $B = 1.5$ used throughout this paper gives nearly the minimum buckling pressure in the range $0 \leq \delta/h \leq 1$ (c.f., [1,2]). This choice also ensures that the imperfection is confined to the pole diminishing to zero at distances on the order of $\sqrt{Rh}$ from the pole.

The underlying assumptions for this theory are small strains, $|\varepsilon| \ll 1$, and moderate rotation, $\varphi^2 \ll 1$. The resultant membrane stresses, $(N_\theta, N_\omega)$, and bending moment quantities, $(M_\theta, M_\omega)$, which are work conjugate to the corresponding strains, have dimensionless forms

$$(n_\theta, n_\omega) = (N_\theta, N_\omega)/(Eh) \text{ and } (m_\theta, m_\omega) = (M_\theta, M_\omega)R/D \tag{2.4}$$

where $D = Eh^3/[12(1-\nu^2)]$ is the bending stiffness. These dimensionless stresses are related to the dimensionless strain quantities by

$$(n_\theta, n_\omega) = (\varepsilon_\theta + \nu\varepsilon_\omega, \varepsilon_\omega + \nu\varepsilon_\theta)/(1-\nu^2), \quad (m_\theta, m_\omega) = (\kappa_\theta + \nu\kappa_\omega, \kappa_\omega + \nu\kappa_\theta) \tag{2.5}$$

In terms of dimensional quantities, the principle of virtual work for axisymmetric behaviour is

$$2\pi R^2 \int_0^{\pi/2} \{M_{\theta\theta}\delta K_{\theta\theta} + M_{\omega\omega}\delta K_{\omega\omega} + N_{\theta\theta}\delta E_{\theta\theta} + N_{\omega\omega}\delta E_{\omega\omega}\}\cos\theta d\theta = \\ -2\pi R^2 \int_0^{\pi/2} p\delta w \cos\theta d\theta + 2\pi R^2 \int_0^{\pi/2} \{f_r\delta w + f_\theta \delta u\}\cos\theta d\theta \tag{2.6}$$

where a positive pressure $p$ acts inward. The D'Alembert 'acceleration forces' are $f_r = -\rho h \partial^2 w/\partial t^2$ and $f_\theta = -\rho h \partial^2 u/\partial t^2$. The equilibrium equations generated by this principle, expressed in terms of the dimensionless quantities, are

$$-\bar{m}_\theta'' - (\bar{m}_\omega \tan\theta)' + \alpha[\bar{n}_\theta + (\bar{n}_\theta\varphi)' + \bar{n}_\omega] = -\bar{p} + \bar{f}_r \tag{2.7}$$

$$-\bar{m}_\theta' - \bar{m}_\omega \tan\theta - \alpha[\bar{n}_\theta' - \bar{n}_\theta\varphi + \bar{n}_\omega \tan\theta] = \bar{f}_\theta \tag{2.8}$$

Here, $(\bar{n}_\theta, \bar{n}_\omega, \bar{m}_\theta, \bar{m}_\omega) = \cos\theta(n_\theta, n_\omega, m_\theta, m_\omega)$, $\alpha = 12(1-\nu^2)(R/t)^2$, $(\ )' = \partial(\ )/\partial\theta$ and

$$\bar{p} = \cos\theta \frac{pR^3}{D}, \quad (\bar{f}_r, \bar{f}_\theta) = -\cos\theta\left(\frac{\partial^2 W}{\partial \tau^2}, \frac{\partial^2 U}{\partial \tau^2}\right) \tag{2.9}$$

with dimensionless time

$$\tau = \sqrt{\frac{D}{\rho h R^4}}\, t \tag{2.10}$$

For solutions symmetric about the equator of the shell, the boundary conditions at the pole and at the equator require $U=0$, $\varphi=0$ and $W'''=0$.

The dimensionless system is defined by the parameters $v$, $R/h$, $\delta/h$ and $\beta_l$. Most of our results will be essentially independent of $R/h$ because the values of this parameter chosen are large enough such that the solutions approach the solution limit for large $R/h$.

*(b) Discretization in space*

Equations (2.7)–(2.9) form a nonlinear system of partial differential equations in space (angle $\theta \in [0, \pi/2]$) and dimensionless time $\tau$. For quasi-static equilibrium computations (figures 1 and 2), this system is solved with zero acceleration forces ($\bar{f}_r = \bar{f}_\theta = 0$) and varying $\bar{p}$ as a free parameter to obtain the curves of equilibria, as shown in figures 1 and 2 below. The functions $\varphi$ and $W$ are approximated by continuous piecewise polynomials of degree 5, consisting of 100 pieces. This piecewise polynomial collocation approximation in the angle $\theta$ is supported by embedded boundary-value solvers for ordinary differential equations (ODEs) such as the collocation toolbox of COCO [11], which was used for figures 1 and 2 below, and similar solvers in previous publications on quasi-static buckling problems [2, 11]. The mesh in $\theta$ is non-uniform: the length of a subinterval is approximately 0.1 close to the equator and approximately $10^{-3}$ near the pole, equi-distributing an error estimate.

*(c) Discretization in time*

For dynamic simulations with a time step size $\Delta_t$ we keep the space ($\theta$) mesh constant over time, and use the BDF-2 rule for approximating time derivatives. BDF-2 approximates the time derivative $\dot{y}(t)$ of function $y(t)$ at time step $t_k = k\Delta_t$ by the finite backward difference $\text{BDF}_2(y)(t_k)$, which depends on the values of $y$ at the current and previous two time steps,

$$\dot{y}(t_k) \approx \text{BDF}_2(y)(t_k) = \frac{1}{\Delta_t}[a_0 y(t_k) + a_1 y(t_{k-1}) + a_2 y(t_{k-2})], \qquad (2.11)$$

where $(a_0, a_1, a_2) = (1.5, -2, 0.5)$ for $k > 1$ and $(a_0, a_1, a_2) = (1, -1, 0)$ for $k = 1$ (see [21]). We used the overdot to denote derivative with respect to dimensionless time $\tau$, and dropped the argument $\theta$ in (2.11). For the dynamic simulation we solve at each time step $t_k = k\Delta_t$ the nonlinear system (2.7)–(2.9) the same way as for equilibrium computations to obtain the solutions $W(t_k)$ and $\varphi(t_k)$. We introduce the additional variables $(W_\tau(t_k), U_\tau(t_k))$ approximating the time derivatives $(\dot{W}(t_k), \dot{U}(t_k))$.

In the (now non-zero) acceleration forces (see eq.(2.9)) we replace the term $\partial^2 W/\partial^2 \tau$ by $\mathrm{BDF}_2(W_\tau)(t_k)$ in $\overline{f}_r$, the term $\partial^2 U/\partial^2 \tau$ by $\mathrm{BDF}_2(U_\tau(t_k))$ in $\overline{f}_\theta$ (both in (2.9), entering (2.7) and (2.8)), and add the equations $W_\tau(t_k) = \mathrm{BDF}_2(W)(t_k)$ and $U_\tau(t_k) = \mathrm{BDF}_2(U)(t_k)$. This results in a closed nonlinear system of equations (as many equations as variables) for the variables $(W(t_k), \varphi(t_k), W_\tau(t_k), U_\tau(t_k))$ at every time step $t_k = k\Delta_t$. This system has the same structure as the problem solved for equilibrium computations. It can, thus, be solved with the same solver. In fact, in the limit $\Delta_t \to \infty$, this nonlinear system for the dynamic simulation approaches the nonlinear system for equilibria. For dynamic simulations with a pressure ramp the variable $\overline{p}$ is a given function of time. The discretization introduces an error of order $(\Delta t)^2$.

*(d) Energy, work and volume*

We conclude this section by listing some fundamental quantities in dimensionless form which will be employed in the paper. Symmetry about the equator allows the expressions below for the full shell to be reduced to integration over the upper half of the shell. At any stage of deformation, the strain energy in the shell, $SE$, is [2]

$$\frac{SE}{2\pi D} = \frac{1}{2}\int_{-\pi/2}^{\pi/2}\left\{\left(\kappa_\theta^2 + 2\nu\kappa_\theta\kappa_\omega + \kappa_\omega^2\right) + 12\left(\frac{R}{h}\right)^2\left(\varepsilon_\theta^2 + 2\nu\varepsilon_\theta\varepsilon_\omega + \varepsilon_\omega^2\right)\right\}\cos\theta d\theta \qquad (2.12)$$

The linearized expression for the decrease of volume of the shell from its unstressed state, $\Delta V$, is sufficiently accurate for this study and is given by

$$\frac{\Delta V}{2\pi R^3} = -\int_{-\pi/2}^{\pi/2} W\cos\theta d\theta \qquad (2.13)$$

The kinetic energy of the shell, $KE$, is

$$\frac{KE}{2\pi D} = \frac{1}{2}\int_{-\pi/2}^{\pi/2}\left\{\left(\frac{\partial W}{\partial \tau}\right)^2 + \left(\frac{\partial U}{\partial \tau}\right)^2\right\}\cos\theta d\theta \qquad (2.14)$$

with the dimensionless time $\tau$ defined in (2.10). The cumulative work done by the pressure, $W_p$, for a shell that is unstressed at $t = 0$ is

$$\frac{W_p}{2\pi D} = -\int_0^\tau d\tau \int_{-\pi/2}^{\pi/2} \bar{p} \frac{\partial W}{\partial \tau} d\theta \qquad (2.15)$$

While the governing equations (2.7)-(2.9) conserve the energy balance $W_p = SE + KE$, the time discretization introduces some loss of energy (damping) that increases with $\Delta t$. In our discussion of the results we will always specify the time step $\Delta t$ and discuss the relation between damping and $\Delta t$ in Section 5.

Finally, we will present results using a second dimensionless time defined by $\hat{\tau} = t/T_0$, where $T_0$ is the period of the sinusoidal spherically symmetric vibration mode of the unpressurized shell, also known as the breathing mode, (correspondingly, $\omega_0 = 2\pi/T_0$ is the breathing frequency) and readily derived as

$$T_0 = \frac{2\pi}{\omega_0} = \sqrt{\frac{2(1-\nu)\rho}{E}} \pi R \qquad (2.16)$$

## 3. Selected results for buckling under uniform pressure relevant to dynamic buckling

As background for the dynamic buckling, we present a brief overview of results for the buckling of a spherical shell subject to a quasi-statically applied uniform inward pressure. Figure 1 reminds the reader of the axisymmetric buckling of a complete spherical shell under spatially uniform external pressure, presented here based on our formulation in Section 2 and using the current notation. In each graph the vertical axis displays the ratio of the pressure to the classical critical buckling pressure, $p_C$, of the perfect shell from the linearized analysis. The horizontal axis in figure 1a is the inward deflection at the pole divided by the shell thickness $h$, while the horizontal axis in figure 1b is the volume decrease normalized by the volume decrease of the perfect shell at the classical buckling pressure, $\Delta V_C$. The results based on moderate rotation theory in Fig. 1 agree closely with results computed independently using FEM modeling in [1]. The classical values are [1]

$$p_C = \frac{2E(h/R)^2}{\sqrt{3(1-v^2)}} \quad \text{and} \quad \Delta V_C = \frac{4\pi(1-v)R^2 h}{\sqrt{3(1-v^2)}} \tag{3.1}$$

Both graphs display the static equilibrium states of imperfect shells with the imperfections indicated. The upper curve for the smallest imperfection, $\delta/h = 1/640$, is a close approximation to the behaviour of the perfect shell. Under the slow quasi-static increase of the controlled (dead) pressure, the buckling pressures (the maximum pressures) are indicated by small black dots. At these limit points (called folds or saddle-node bifurcations in dynamics) the imperfect shell in a noise-free environment will jump dynamically (snap buckle) to a collapsed state outside the range of this theory.

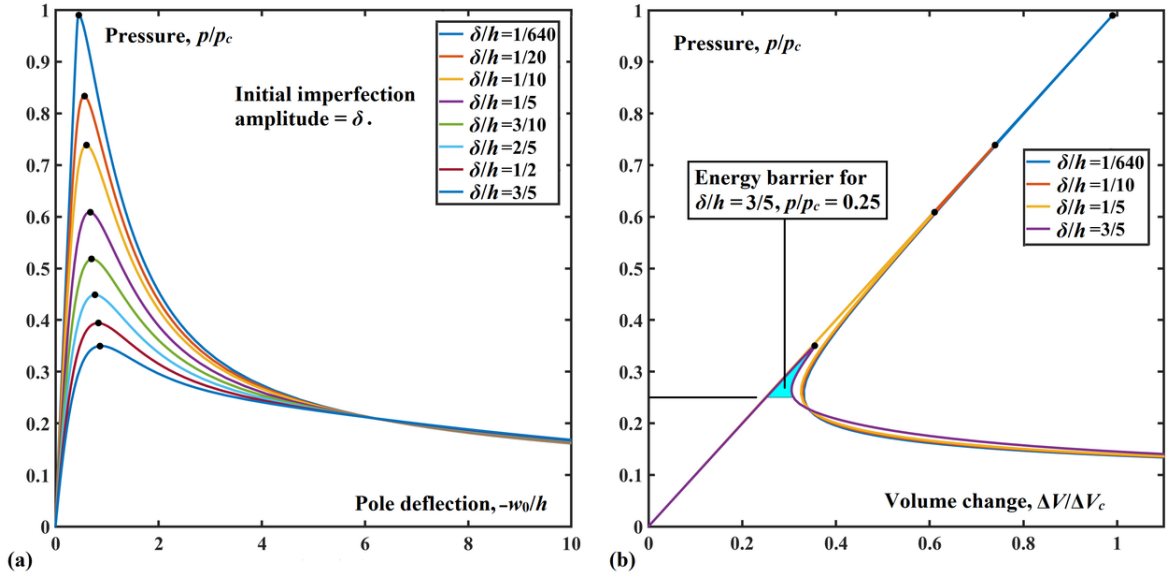

Figure. 1 Buckling behaviour of spherical shells with $R/h = 100$, $v = 0.3$ for various dimple imperfections having $B = 1.5$. (a) Pressure versus pole deflection. (b) Pressure versus change in volume with the energy barrier to buckling illustrated for a prescribed pressure, $p/p_C = 0.25$, for the shell with $\delta/h = 3/5$.

As the load increases slowly towards these limit points, a shell is in a more and more precarious meta-stable state protected by a diminishing energy barrier against small disturbances (e.g., noise) typically present in a physical experiment. The magnitude of this barrier $E_{Barrier}$ of a given shell at a prescribed (fixed) pressure $p$ can be identified as the 'triangular- shaped' area on the plot of $p(\Delta V)$ in figure 1b. This area is simply the difference of the energy of the shell/loading system between the unstable buckled state on the falling segment of the curve (a dynamicist's saddle) and the stable un-buckled state on the rising segment (a dynamicist's node). This area is the difference in the strain energy of the shell in the two states less the work $p\Delta V$ that would be performed by the external pressure.

Energy barriers have been accurately calculated in [6, 12] for perfect and imperfect spherical shells under both (dead) controlled pressure and (rigid) controlled volume loading conditions. The barrier for prescribed pressure has been recomputed and presented in figure 2 using the same imperfection amplitudes employed in figure 1. For prescribed pressure, the quasi-static system energy is $\psi = SE - p\Delta V$. The results shown in figure 2 for various levels of imperfection are independent of $R/h$, to a very good approximation for thin shells with $R/h \geq 50$. The energy barrier vanishes at the buckling pressure and remains very small for pressures or volume changes somewhat below the buckling value. At pressures well below the buckling value the energy barrier increases dramatically and becomes relatively insensitive to the imperfection level. The energy barrier in figure 2 is normalized by $\frac{1}{2}p_C \Delta V_C Ch/R$ where $C = \sqrt{3}/\left[(1-\nu)\sqrt{1-\nu^2}\right]$. Note that $\frac{1}{2}p_C \Delta V_C$ is the strain energy in the perfect shell at the classical buckling pressure. It follows from figure 2 that, because of the factor $Ch/R$, the energy barrier is a very small fraction of the energy stored in a thin shell (or, equivalently, of the work done on the shell by the pressure). Moreover, the ratio of the energy barrier to the stored energy decreases for thinner shells directly in proportion to $h/R$. This is due to the fact that the deformation in the buckled state is localized at the pole in the form of a dimple whose size scales with $\sqrt{Rh}$ and thus decreases in size relative to the shell itself as $h/R$ diminishes, as will be discussed more fully later. The implication of this will be discussed later with regard to dynamic buckling under step loading.

In the limit for very thin shells, $h/R \to 0$, the energy barriers for prescribed pressure and prescribed volume change are the same. The barrier in the case of prescribed volume change does depend somewhat on $h/R$, as discussed in [11].

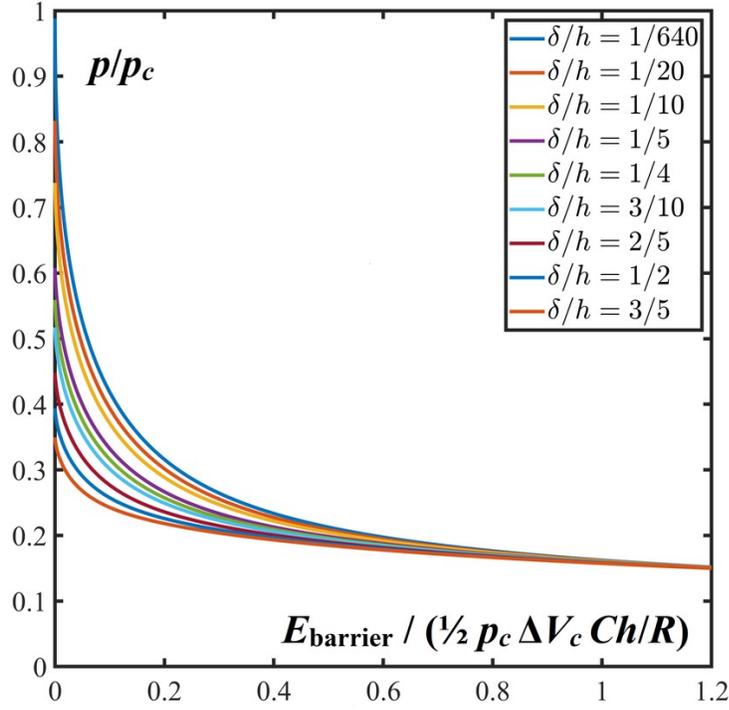

Figure. 2 Energy barrier, $E_{barrier}$, for shells loaded to prescribed dead pressure $p$. The barrier is presented for various levels of imperfection; these results are essentially independent of $R/h$ for $R/h$ greater than about 50. For these results, $R/h = 100$, $\nu = 0.3$ and $B = 1.5$.

We conclude this section a few additional results relevant to dynamic buckling. The vibration frequency of the breathing mode of the perfect shell ($\delta/h = 0$) undergoing small spherically symmetric oscillations ($w'(\theta) = 0, u(\theta) = 0$) and subject to no applied pressure introduced earlier in (2.8) is (equivalent to (2.16)

$$\omega_0 = \sqrt{\frac{2E}{(1-\nu)\rho}} \frac{1}{R} \tag{3.2}$$

The buckling mode of the perfect shell associated with the lowest eigenvalue of the classical buckling problem, $p_C$, is also an important reference mode. The lowest vibration frequency of the perfect shell vanishes at $p_C$ and the associated mode is the classical buckling mode. The normal deflection of the classical mode can be expressed in terms of a Legendre polynomial—explicit expressions are given in [2,13].

The spectrum of frequencies of the modes linearized about applied pressures below $p_C$ is also revealing and relevant to the understanding of dynamic buckling. An illustration is presented in figure 3 which shows a selection of normalized complex frequencies, $\omega/\omega_0$, and associated modes for a shell with imperfection amplitude, $\delta/h = 1/4$, and subject to pressure $p = 0.512 p_C$. Note from figure 1 that there are two equilibrium solutions associated with $p = 0.512 p_C$, one the stable unbuckled state (written here as 'node') where the pole deflection is approximately $w_0 = -0.4h$ and the other the unstable equilibrium point at the saddle point of the system energy (written as 'saddle') where the pole deflection is approximately $w_0 = -1.1h$. The results in figure 3 are obtained by linearizing the equations about these two equilibrium solutions. The time dependence of the linearized solution has the form $e^{\omega t}$ where $\omega$ is the complex frequency for the respective mode. Note that with normalization used in figure 3a, the reference breathing mode for the perfect unpressured shell has $\omega/\omega_0 = \pm i$.

The spectrum of frequencies in the two states at $p = 0.512 p_C$ are plotted in figure 3a and two of the most important associated modes shapes for each state are plotted in figure 3b. Since the shell has an imperfection there is no strictly spherically symmetric motion, but the mode identified as the breathing mode in the unbuckled state, which has $\omega/\omega_0$ nearly equal to $\pm i$ (pair of black dots in figure 3a identified by A). The associated normal deflection of mode A is plotted in the upper half of Fig. 3b deviates from the uniform normal deflection of the breathing mode of the perfect shell due to the imperfection. If one tracks this 'breathing' mode through the equilibrium solutions to the buckled state at $p = 0.512 p_C$, one finds that the normalized frequencies hardly change from $\pm i$ ( pair of black dots labeled D in the saddle spectrum), and the associated mode has even more distinct $\theta$-variations associated with the non-uniformity of the buckled state about which the linearized solution has been obtained. Next, we focus on the lowest vibration frequency of the unbuckled shell at $p = 0.512 p_C$ for the mode identified as the 'buckling' mode, which as seen by the second set of dots labeled C in Fig. 3a has $\omega/\omega_0 \cong \pm 0.35i$. The normal deflection of the associated mode plotted in figure 3b is very similar to that of the classical buckling mode. Tracking this mode to the unstable buckled state at $p = 0.512 p_C$ leads to the 'buckling mode' with $\omega/\omega_0 \cong \pm 0.34$ in figure 3a (labelled B) corresponding to exponential growth/decay. An important feature regarding the buckling mode shape is that the undulations have localized to the polar region in the form of a dimple.

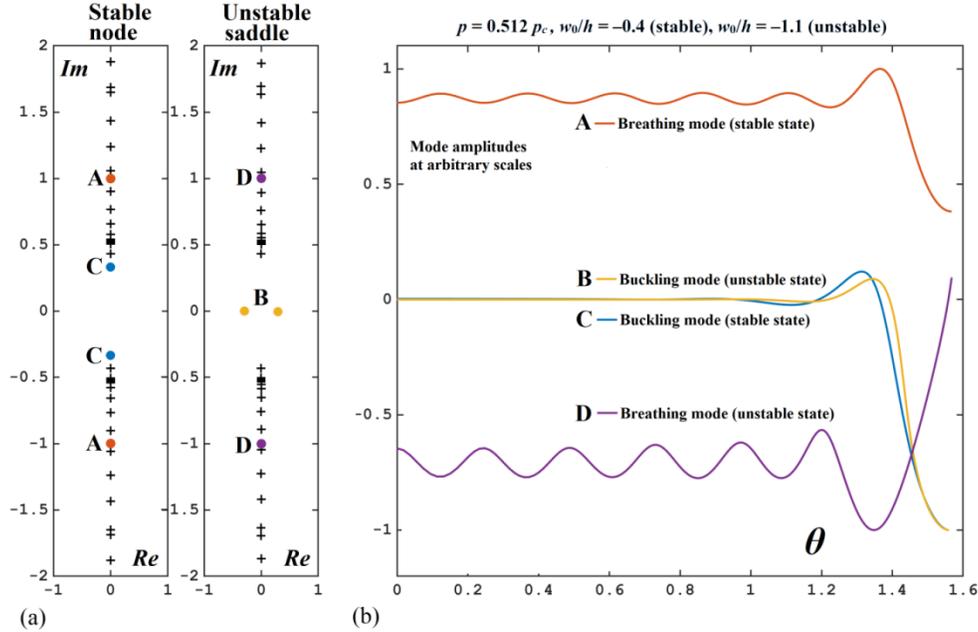

Figure. 3 Linearized modes about the two equilibrium states at $p = 0.512 p_C$, the stable unbuckled state (node) and the unstable buckled state (saddle). The shell has imperfection, $\delta/h = 1/4$ and $B = 1.5$, with $R/h = 100$ and $\nu = 0.3$. The normalized spectrum of the frequencies, $\omega/\omega_0$, is given for the two states in a) and the radial displacements of two of the most important modes are presented in b). The frequencies and their associated modes are identified and discussed in the text. All modes have been scaled to have maximum modulus 1.

## 4. Buckling under step loading with spatially uniform pressure

As noted, attention in this paper is limited to the response of spherical shells under pressure loads that are spatially uniform. In subsequent sections we will consider dynamic buckling under time varying spatially uniform pressure, $p(t)$, in which the shell starts at rest in an unpressured state. From the stationary starting state, the pressure is ramped up rapidly and then held constant at its end value, which we also call $p$ (without argument $t$) in our discussions.

We refer to this loading as step loading. For the limit in which ramping is instantaneous, we use the terminology 'instantaneous step loading'.

Figure 4 shows the dynamic response of several quantities of interest for a step loaded imperfect shell with $\delta/h = 0.25$ such that its quasi-static buckling pressure is $p_S(\delta) = 0.57 p_C$. The pressure is ramped up rapidly from 0 at $t/T_0 = 1$ to $p = 0.512 p_C$ at $t/T_0 = 1.25$ (about 10% below the static buckling pressure) and held at $0.512 p_C$ for $t/T_0 > 1.25$. The plots in figure 4a show the time-variation of the work done by the pressure, $W_p/2\pi D$, the strain energy in the shell, $SE/2\pi D$, and the kinetic energy, $KE/2\pi D$. In the simulation run for figure 4 the time step size $\Delta t$ is chosen sufficiently small such that the energy balance, $(SE + KE - W_p)/2\pi D = 0$, is satisfied to a high degree of accuracy (see Section 5 for a detailed discussion of damping).

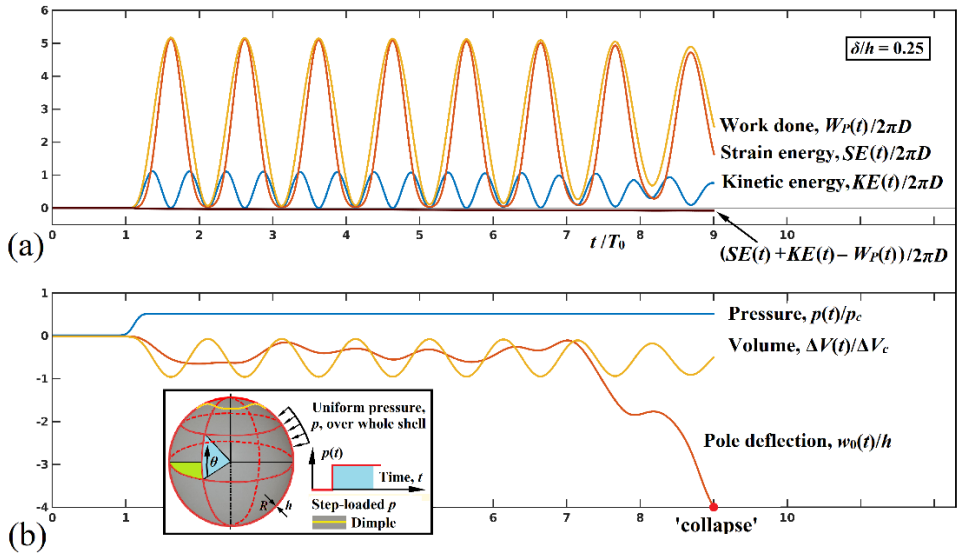

Figure. 4 Time profiles of applied pressure, $p/p_C$, together with the cumulative work of pressure, $W_p/2\pi D$, strain energy, $SE/2\pi D$, kinetic energy, $KE/2\pi D$, volume change, $\Delta V/\Delta V_C$, and pole deflection, $w_0/h$. These are for an imperfect shell with imperfection amplitude $\delta = 0.25h$, The other parameters are $R/h = 100$, $\nu = 0.3$, $B = 1.5$ with time step size $\Delta_t = 0.022$. The static buckling pressure is $0.57 p_C$ and the steady dynamic pressure in these simulations is $0.512 p_C$. The inset shows the geometry underlying Eqs. (2.1-2.9).

The lower plots in figure 4b show the prescribed time variation for the pressure and the variation of the volume change, $\Delta V / \Delta V_C$, and the pole deflection, $w_0 / h$. Note that for seven cycles the strain energy, work done by the pressure and the volume change are nearly sinusoidal with period $T_0$. During these cycles the kinetic energy has a period $T_0 / 2$ as is typical for vibratory systems. Prior to $t / T_0 \cong 7$ the motion is dominantly a spherical symmetric 'breathing' oscillation with $(w', u) = 0$. However, from the beginning, the deflection at the pole responds differently from most of the shell and, for this example, starting at roughly $t / T_0 \cong 7$ conditions at the pole give rise to localized snapping into a dimple buckle. In all cases in this paper, an inward pole deflection exceeding 3 or 4 time the thickness results in buckling of the spherical shell. Once the pole deflection reaches this magnitude the shell cannot resist dynamic collapse. The buckling behaviour is brought out more clearly in figure 5 which displays two representations of the normal deflection of the shell and one representation of the in-plane displacement as functions of both position $\theta$ and time. By $t / T_0 = 8$ it is evident that the deflection has taken the form of a dimple localized at the pole surrounded by spatial undulations decaying away from the pole. The amplitude of the dimple at the pole doubles between $t / T_0 = 8$ and $t / T_0 = 9$. Under the fixed pressure the dimple grows unabated leading to complete collapse of the shell. The in-plane displacement is roughly two orders of magnitude smaller than the normal deflection, which is typical for spherical shell buckling.

Two aspects of the dynamic process stand out from figures 4 and 5: the significant delay in the formation of a buckle until about 6 or 7 overall oscillations of the shell in this particular case, and the localization of the buckle at the pole as it develops. Further insight into the delay in buckling will emerge when results are presented shortly for the responses of the shell to a full range of step pressures. The localization helps to explain why the various energy variations of figure 4 are very large compared to the energy barrier to buckling, and it will be useful at this point to highlight these energy differences. Note that for the example in figure 4 the cyclic variations in the strain energy and kinetic of the shell prior to buckling are $SE / 2\pi D \cong 5$ and $KE / 2\pi D \cong 1$. At a pressure $p / p_C = 0.512$, the energy barrier separating the static un-buckled and buckled states for an imperfect shell with $\delta / h = 0.25$ can be obtained from figure 2 as $E_{Barrier} / [\frac{1}{2} p_C \Delta V_C Ch / R] \cong 0.00436$. The conversion between the two normalization factors in these dimensionless energy ratios is

$$\frac{1}{2} p_C \Delta V_C \frac{Ch}{R} = \frac{8\sqrt{3}}{\sqrt{1-v^2}} \frac{h}{R} (2\pi D) \qquad (4.1)$$

For the shell with $R/h = 100$ and $v = 0.3$, the conversion factor is $8\sqrt{3}h/[(1-v^2)R] = 0.145$ and thus $E_{Barrier} / 2\pi D \cong 0.000633$. The energy barrier is a tiny fraction of the energy variations taking place in the shell. Apart from one aspect mentioned later related to the level of imperfection, the barrier has essentially no quantitative relevance to the uniform pressure step loading because only a small region of the shell near the pole participates in the buckling process. Most of the shell undergoes breathing motion with $(w', u) \cong 0$ which accounts for nearly all of the energy variations seen in figure 4. The coupling between the breathing motion and the emerging dimple buckle at the pole requires seven cycles before buckling occurs.

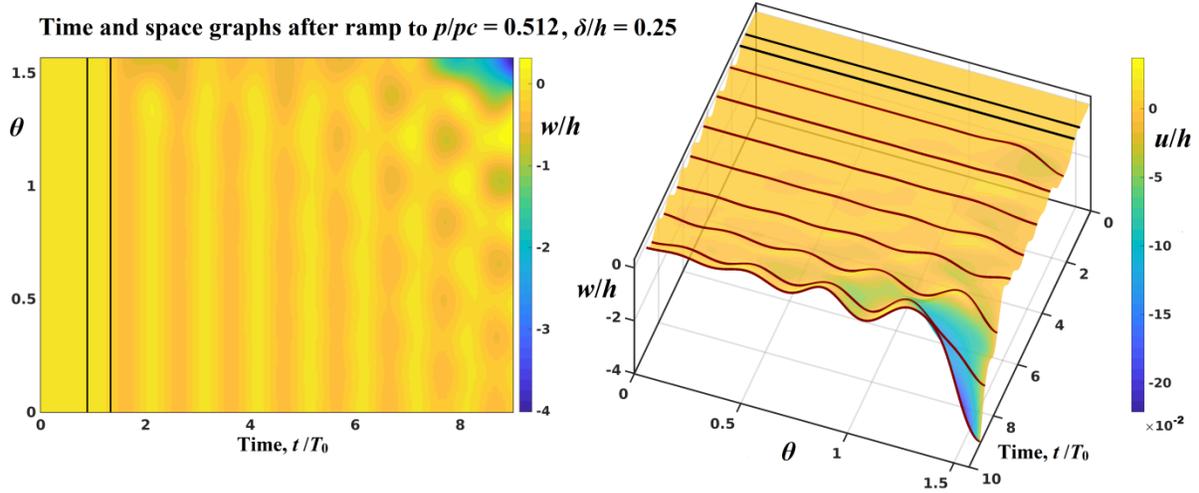

Figure 5. Normal deflection, $w(\theta, t)/h$, (with the in-plane displacement, $u(\theta, t)/h$, included on the right) plotted as a function of position, $\theta$ in radians, and time, $t/T_0$, for the step loading example presented in figure 4.

The dynamic responses at the pole of an imperfect shell with $\delta/h = 0.1$ and subject to various levels of step loading is shown in Fig. 6a revealing whether buckling takes place or not.

The companion plot in figure 6b, perhaps the most important figure in this paper, summarizes the dynamic buckling behaviour under step loading over the range of imperfections from $\delta/h=0$ to $\delta/h=0.6$. We systematically ran a sequence of simulations for a range of amplitudes $\delta$ (from $h/640$ to $0.6h$) and a range of rapid uniform pressure ramps from 0 up to a final pressure $p$. Each dot in the right panel of figure 6 corresponds to one simulation: a dot at $(\delta, p)$ means that a simulation was run with pressure increasing from 0 to $p$ between $t/T_0 = 1$ and 1.5 and then held constant up to time $t/T_0 = 26$. If the pole deflection dropped below $-4h$ at a time $t_b/T_0$ before the end time, we record the simulation as "buckled", colouring the dot red and indicate the delay $(t_b - 1.5)/T_0$ in the contours. Otherwise, the dot is coloured green indicating that the simulation did not show buckling. The left panel of figure 6 shows time profiles of the pole deflection $w_0(t)$ for a sequence of simulations for a fixed imperfection amplitude $\delta/h=0.1$. Time profiles in green colour did not buckle before $t/T_0 = 26$, those in dark red did buckle. The transparent surface shows the static saddle equilibrium value for the pole deflection (with a small part of the node equilibrium surface close to the fold value $p_S = 0.74 p_C$). Blue ellipses indicate where the time profiles cross this saddle surface. All time profiles that cross the saddle surface wound up buckling.

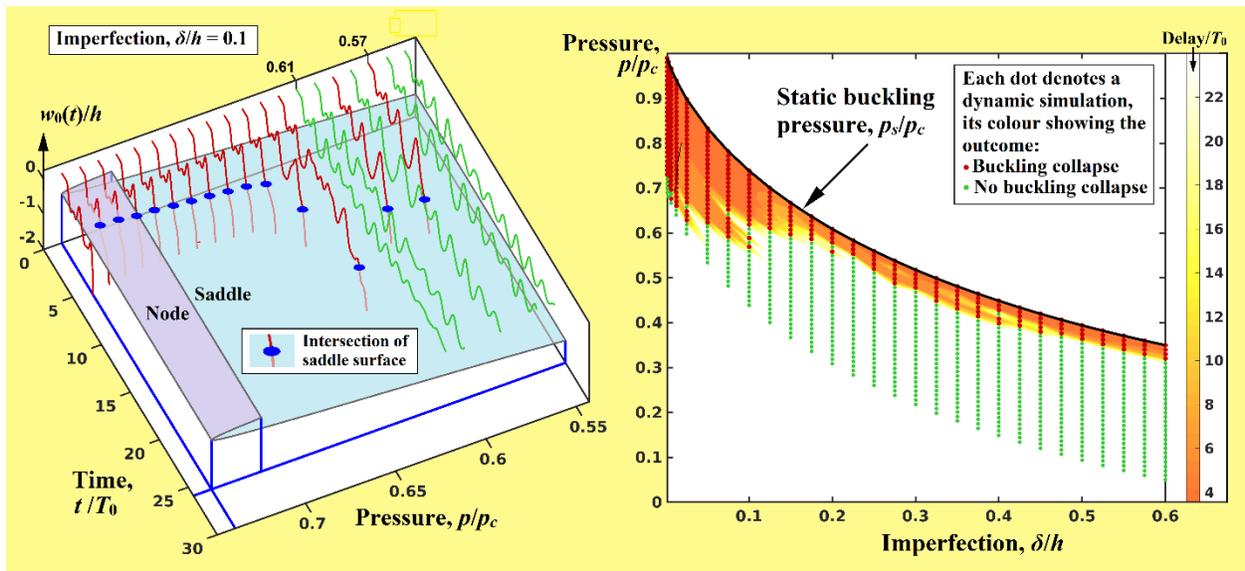

Figure 6: Parameter scan over end pressures $p$ for pressure ramp times of $T_0/2$ and dynamic buckling summary for imperfection amplitudes in the range $0 \leq \delta/h \leq 0.6$. Each dynamic

simulation is carried out from $1 \leq t/T_0 \leq 26$. Left panel: Sequence of time profiles for pole deflections $w_0(t)$ for fixed $\delta/h = 0.1$ and varying end pressure $p$ (21 evenly spaced values between $0.54 p_C$ and $0.74 p_C = p_S$). Right panel: Overview of simulation results in the ($p$, $\delta$) plane showing parameter values where buckling occured (red dots) or did not occur before $t/T_0 = 26$ (green dots). The static buckling pressure $p_S(\delta)$ is the black curve. The colour encodes the delay of the buckling after the end of the pressure ramp in units of $T_0$. Parameters: $R/h = 100$, $\nu = 0.3$, $B = 1.5$, $\Delta_t = 0.087$.

Figure 6 permits two observations. First, there is not a clear-cut buckling threshold for step loading. In the left panel a ramp up to a final pressure of $0.57 p_c$ leads to buckling, while a ramp up to a somewhat larger final pressure of $0.61 p_c$ does not give rise to buckling. Second, for imperfections that are not too small (for $\delta/h \geq 0.15$), the difference between the static buckling pressure and the lowest step pressure causing buckling is uniformly small (about $0.05 p_c$), while for small imperfections (e.g., $\delta/h < 0.15$) the difference is significantly larger, with reductions up to $0.3 p_c$ (when $\delta/h \cong 0$). *Especially for shells with small imperfections, the static buckling pressure is not an accurate estimate of the buckling pressure for a step pressure loading.*

This finding is at odds with the dynamic buckling predictions for step loading based on 1-degree-of-freedom (DOF) models analyzed in [14,15]. These authors considered two types of imperfection-sensitive, 1-DOF models: one with unstable symmetric bifurcation behaviour (with a cubic nonlinearity) and the other with asymmetric bifurcation behaviour (with a quadratic nonlinearity). In each case, for every level of imperfection, it was possible to relate the instantaneous step buckling load, $\lambda_D$ (generalizing $p_D$), to the static buckling load, $\lambda_S$ (generalizing $p_S$), and the static buckling load of the perfect model, $\lambda_C$ (generalizing $p_C$). For the symmetric model, this relation is

$$\frac{\lambda_D}{\lambda_S} = \frac{1}{\sqrt{2}} \left( \frac{\lambda_C - \lambda_D}{\lambda_C - \lambda_S} \right)^{3/2} \tag{4.2}$$

while for the asymmetric model it is

$$\frac{\lambda_D}{\lambda_S} = \frac{3}{4}\left(\frac{\lambda_C - \lambda_D}{\lambda_C - \lambda_S}\right)^2 \qquad (4.3)$$

The relations between the step buckling load and the static buckling load from the 1-DOF models discussed above are in complete agreement with asymptotic results obtained by Thompson [16] derived from a general ($n$- DOF) formulation for discrete elastic systems for which the perfect system has a unique buckling mode. The analysis is purely quasi-static and asymptotic for small imperfections, but did compare well with step-buckling experiments on structural frames of the type built by J. Roorda [24]. For both unstable symmetric and asymmetric bifurcations, Thompson determined: 1) the relation between the static buckling load $\lambda_S$ (the max load) and the imperfection, and 2) the relation between the 'astatic load', $\lambda_N$, and the imperfection. For a given imperfection, the astatic load is that load $\lambda$ at which the work done by the fixed $\lambda$ equals the strain energy in the unstable buckled state. The asymptotic results for the astatic load coincide with asymptotic limits of (4.2) and (4.3) for small imperfections if $\lambda_N$ is identified with $\lambda_D$. For instantaneous step loading of the 1-DOF models discussed above, it is straightforward to prove that the astatic load $\lambda_N$ must be a lower bound to the dynamic buckling load $\lambda_D$. The fact that $\lambda_N = \lambda_D$ for these models is due to the fact that, at the lowest step load for which the model buckles, the model attains the static unstable buckled state, momentarily coming to rest, such that the astatic condition is exactly satisfied.

Applied to our spherical shell, the astatic condition is easily visualized as shown in figure 7. When the two areas, *Area*1 and *Area*2, on the pressure-volume plot in figure 7 are equal, the astatic condition for $p_N$ is met, i.e., $p_N \Delta V_B = SE_B$ with $B$ denoting the unstable static buckled state. At pressures below $p_N$, the strain energy in state $B$ exceeds $p_N \Delta V_B$, and vice versa. Thus, if an instantaneous step-load occurred to a pressure lower than $p_N$, the loading system would not be able to provide enough work to reach the saddle point represented by $B$. Although we have not proved that the astatic pressure, $p_N$, is a lower bound to the instantaneous step buckling pressure, it seems likely that this is the case. For $p_N$ to qualify as a *strict* lower bound (even an excessively conservative one) one would have to establish that there are no other mountain passes associated with other unidentified saddle equilibria. However, the detailed investigation of the thresholds later in Section 6 indicate that escape occurs indeed near the saddle used in computing the energy barrier presented in figure 2 and discussed further in connection with figure 7.

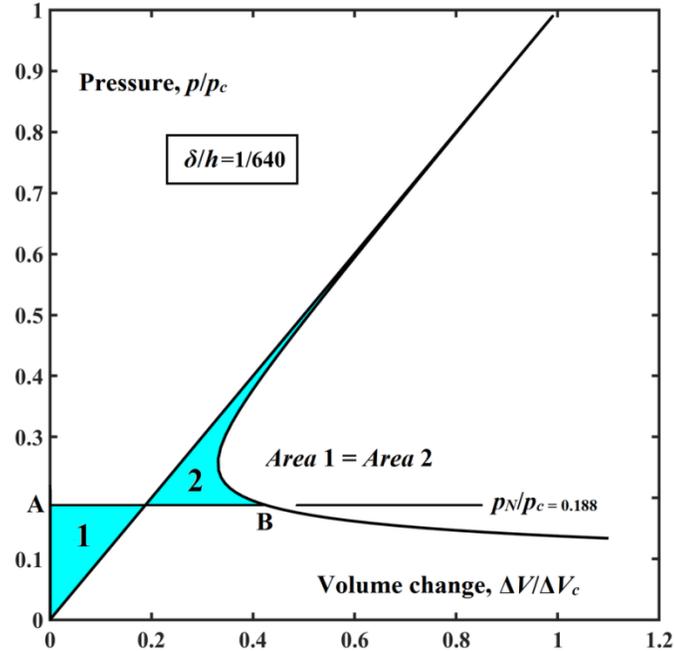

Figure. 7 Illustration of equal area construction for determining the astatic pressure $p_N$ based on the condition $p_N \Delta V_B = SE_B$. The case illustrated for $\delta/h = 1/640$ has $p_N / p_C = 0.188$.

We have computed the normalized astatic pressure $p_N / p_C$ as a function of the imperfection amplitude for the shells considered in figure 1. We find that the astatic pressure is $p_N / p_C \approx 0.2$ with almost no dependence on the imperfection amplitude for imperfections in the range $\delta/h \leq 0.6$. Compared with the step buckling pressures in figure 6, the astatic pressure is unrealistically low and of little predictive value, at least for the damping level associated with the results in figure 6. Hoff and Bruce [17] made an early use of the astatic load in their study of the buckling of shallow arches subject to step loading of a pressure distributed along its length. The shallow arch is like the spherical shell in that it undergoes dramatic changes in deflection when buckling occurs—so called snap buckling. It differs in that the entire arch buckles while the buckling deflections of the sphere are localized near the pole. In the one specific example Hoff and Bruce considered, they found the dynamic step loading prediction agreed quite well with the astatic pressure, both giving an estimate of the dynamic buckling pressure that was about 20% below the static buckling pressure. This result is similar to what one might expect based on the 1-DOF models discussed above and on similar models in the book on dynamic stability [7].

The relation between the dynamic and static buckling loads for the spherical shell is at odds with the corresponding relations for the simple 1-DOF models in equations (4.2) and (4.3) as seen in figure 8. For the shell, the largest reductions of the dynamic step buckling pressure relative to the static buckling pressure occur for the shells with the smallest imperfections. By contrast, the dynamic buckling load of the simple models is only slightly reduced from the static buckling load when the imperfection is small. For the models, the largest relative reductions occur for the largest imperfections, while for the spherical shell the opposite is true.

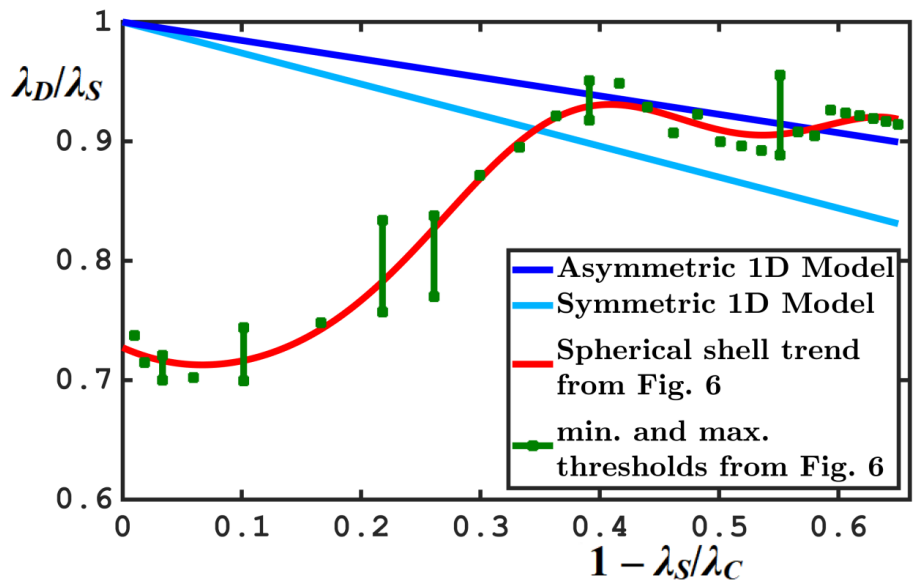

Figure. 8 Ratio of the dynamic buckling load under step loading to the static buckling load plotted as a function of the fractional reduction of the static buckling load from the static buckling load of the perfect structure. Predictions of the 1-degree of freedom models: symmetric model from (4.2) and asymmetric model from (4.3). The trend line from figure 6 for the buckling of the spherical shell under spatially uniform pressure is shown.

The main factor at play in the different dynamic buckling behaviours of the simple models and the spherical shell is associated with the interaction among the different modes activated in the step-loaded shell. The initial response of the sphere is dominated by the oscillatory motion of the breathing mode which absorbs most of the work done by the pressure. When the step pressure is sufficiently large the nonlinear coupling between the breathing mode and the incipient dimple causes the dimple to grow and to snap buckle leading to full collapse of the shell (under the prescribed pressure considered here). The simple 1-DOF models discussed earlier do not encompass modal interaction. Dating from the early work of Goodier and McIvor [18] on the buckling of long cylindrical shells (effectively rings) under dynamic radial pressure there has been a large literature on the coupling of breathing and buckling modes, often leading to a Mathieu equation governing the early stages of the coupling. The ring buckling problem considered in [18] has this form but it is not imperfection-sensitive and the nonlinearity is such that snap buckling does not occur. Instead, in their problem the nonlinear mode interaction gives rise to a gradual amplification of the buckling mode.

Tamura and Babcock [19] carried out an early nonlinear mode interaction analysis for step loading of a finite length, imperfect cylindrical shell under an axial step load. This structure/loading combination is imperfection-sensitive. The oscillation of the axial compressive stress (the breathing mode in this case) excited by the step load was treated approximately and coupled to two interacting buckling modes. The authors analyzed only one specific imperfect shell for which the dynamic buckling load associated with an abrupt increase in the shell deflection was found to be approximately 60% of the static buckling load. More recently, the dynamic buckling of conical shells under step loaded axial compression has been investigated [20]. This problem also has features in common with the spherical shell problem in that the structure/loading system is imperfection-sensitive and results in snap buckling once the buckling mode is sufficiently amplified. In plots of the ratio of the step buckling load to the static buckling load as a function of the imperfection level, results [20] show a trend similar to that in figure 8 for the spherical shell. In particular, they find that conical shells with small imperfections have ratios of dynamic to static buckling as low as about 0.6 and that this ratio increases for larger imperfections, similar to the trend in figure 8. The authors of [20] suggest that their numerical results apply to conditions where damping is negligible, and they do not identify the cascade of buckling thresholds of figure 6. We will return to these issues in the next section.

The plots of the energy barrier for the spherical shell in figure 2 also shed some light on the trend for the dynamic buckling pressure for the spherical shell in Fig. 8. Note that for small imperfections the energy barrier remains very small for values of $p$ as low as 60 to 70% of the static buckling load whereas for larger imperfections the energy barrier remains small for smaller reductions of $p$ relative to the static buckling load. This is consistent with figure 8: a relatively perfect shell is more susceptible to buckling at pressures within a given fraction of its static buckling load than more imperfect shells loaded to the same fraction of their static buckling load.

When snap buckling requires several oscillations of the breathing mode of the shell, as illustrated in figures 4-6, it is obvious that damping effects will influence the dynamic buckling load. Damping is present for these results associated with the numerical algorithm used in solving the dynamic equations. Section 5 which follows discusses some of the issues related to this algorithm and the role of damping in dynamic buckling.

## 5. Balance between damping and nonlinear coupling between modes

Even though the small strain moderate rotation theory does not include any dissipation, some damping is introduced by the numerical BDF-2 time stepper (2.11) for the simulation. As the spectrum of equilibria in figure 3 suggests, the introduction of damping is necessary to make dynamic simulations numerically feasible. Without damping small disturbances of equilibria will lead to near quasi-periodic behaviour composed of oscillations with arbitrarily high frequencies, where the range of frequencies is determined by the resolution of the space discretization. The importance of damping in regularizing the numerical analysis of dynamic structural systems features prominently in modern treatments of the subject [21].

In particular, the introduction of damping has a strong effect on the long-time behaviour of a conservative system such as (2.7)–(2.9). To provide a good estimate of this effect on buckling thresholds, we recap briefly how much damping a time stepper based on the BDF-2 approximation introduces. We also illustrate this effect in figure 9.

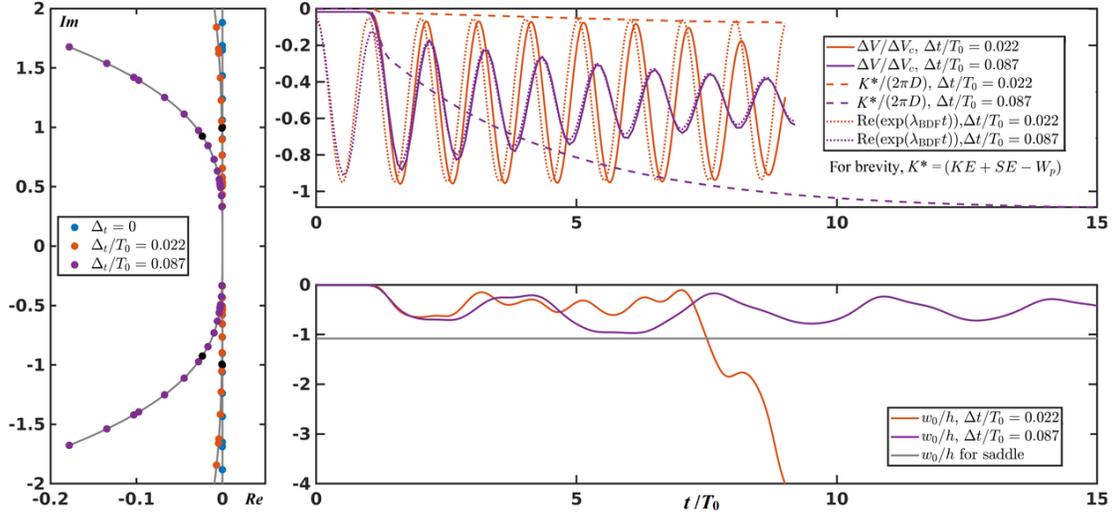

Figure. 9 (Left) Change of spectrum near stable equilibrium (node) from original conservative case (blue) by numerical discretization with small and larger time stepsize. (Right top) Volume oscillations and loss of energy $KE + SE - W_p$ for small and larger stepsize. (Right bottom) Pole deflection for small and larger stepsize. Other parameters: $\delta/h = 0.25$, $p/p_c = 0.512$.

### *(a) Damping at linear level*

The amount of damping a numerical scheme introduces is well understood only for linear systems. In this case one may study the behaviour of the time stepper for each linear mode separately, since the damping depends only on its frequency $\omega$. The damping at frequency $\omega$ is determined by inserting the numerical approximation (in our case the BDF-2 (2.11)) into the linear ODE $\dot{y} = i\omega y$.

The solution of $\dot{z} = i\omega z$ ($\omega > 0$) asymptotes to $z(t) = \exp(\lambda_{\text{BDF}} t)$, where

$$\lambda_{\text{BDF}} = i\omega + d + i\Delta_\omega = \frac{1}{\Delta_t} \log \frac{2 + \sqrt{1 + 2i\omega\Delta_t}}{3 - 2i\omega\Delta_t} \qquad (5.1)$$

Both the numerical growth rate $d$ and frequency shift $\Delta_\omega$ are negative for time step size $\Delta_t > 0$ (going to zero for $\Delta_t \to 0$) such that the time stepper introduces artificial numerical damping $-d(\omega, \Delta_t)$ and a slow-down per period $-\Delta_\omega/(2\pi)$ for a mode with frequency $\omega$. The damping $-d(\omega, \Delta_t)$ can be approximated over the range of frequencies shown in the spectra in figure 3 (to roughly twice the breathing frequency) by expanding the real part of (5.1) in $\omega = 0$:

$$-d(\omega, \Delta_t) = \frac{\omega^4 \Delta_t^3}{4 + 10\omega^2 \Delta_t^2} \quad (5.2)$$

an approximate quartic in the frequency $\omega$. Figure 9 shows the effect and the amount of damping for two different step sizes. The smaller stepsize, $\Delta_t = 0.022 T_0$, was used for the single example trajectories shown in figures 4 and 5, the larger stepsize, $\Delta_t = 0.087 T_0$ (four times the smaller stepsize) was used for the parameter study in Figure 6. Otherwise, all parameters are identical to figures 3, 4 and 5. The left panel of figure 9 shows that the numerical scheme introduces frequency dependent damping, suppressing high-frequency oscillations more strongly, according to the approximately quartic frequency-damping relationship (5.2). Thus, a single small-amplitude breathing oscillation around the stable equilibrium gets damped by less than 0.4% for $\Delta_t = 0.022 T_0$ but by 13.6% for $\Delta_t = 0.087 T_0$ (in one unit of time by our scaling).

*(b) Damping of shell motion after pressure ramp*

The top right of figure 9 shows that the damping factors derived for small-amplitude breathing oscillations carry over to the motion of the shell after the pressure ramp as in figure 4. The volume oscillations are dominated by breathing oscillations and the decay rate of these larger scale breathing oscillations matches the predictions from the linear approximation $\exp(\lambda_{\text{bdf}} t)$ (dotted curves in figure 9, top right panel). The dashed curves show the loss of energy along the trajectory, which is also 16 times higher for the large stepsize $\Delta_t = 0.087 T_0$. We also observe an additional loss of energy during the rapid ramp-up of the pressure in the time period from $T_0$ to $2T_0$, which is not directly predictable from linear theory. However, this loss of energy is, consistently, also higher for the larger stepsize.

*(c) Conclusion for calibration of damping*

Real shells and other numerical discretizations may have damping with frequency dependence different from the one shown in figure 1 and approximated by expansion (5.2). Figure 9 suggests that in these cases damping should be compared at the breathing frequency. In experiments the damping of the breathing vibration can be measured by applying small pressure load ramps far from buckling pressure. According to figure 9 this linear damping carries over to the motion with larger amplitude. Damping in structures has many possible sources, including air resistance, dissipation at joints and boundaries and material damping of various kinds. While the damping generated by the numerical discretization used in the present study may not represent all the physical sources of damping, it is clear from figure 9 and Equation (5.1) that the damping in results from figures 4-6 is comparable to damping in other numerical schemes and empirical data (after calibration at the breathing frequency).

The bottom right panel of figure 9, showing the motion of the pole deflection for both stepsizes (pole deflection of the saddle is shown for reference), demonstrates that the larger damping for the larger stepsize causes the shell to avoid buckling (while it does buckle for lower damping at $\Delta_t = 0.0022 T_0$ as shown also in figures 4 and 5). We expect this to hold in general—lower damping implies lower buckling threshold. One argument for this is given in the next section.

## 6. A cascade of buckling thresholds for non-zero damping

For the nonlinear dynamicist the study in this paper raises a number of interesting fundamental questions, and points to their relevance in practical applications. To examine some of the issues, let us focus on a conservative autonomous system, as is our spherical shell after the pressure has been step-loaded to a fixed value. Additionally, assume that there is only a single potential energy saddle and barrier-height $V_S$ that is preventing escape to a 'remote' region of phase space (such a single saddle is not rigorously established for our shell buckling). This might be thought of as a well understood problem, but this is not the case, especially because our system has many degrees of freedom: strictly an infinite number, but we will treat the shell for simplicity as if it has a large number of mechanical degrees of freedom with a phase-space of N dimensions (twice the DOF number). Within these restrictions, there is a lower bound for both damped and undamped systems because a trajectory starting with total (kinetic plus potential) energy, $E$, cannot escape if $E < V_S$. The question that remains is what happens if $E > V_S$ and the situation is remarkably obscure. Even in the extreme case of no damping and with the elapsed time going to infinity, there is no guaranteed escape due to a number of complex blocking actions. These are still being explored in the multi-body problems of astronomy and chemical kinetics.

The systematic parameter study in figure 6 shows that buckling under step loading can occur in ranges of pressures that are far below the static buckling pressure $p_s$, but also far above lower bounds given by energy barrier. For some imperfection levels $\delta$ multiple buckling thresholds are visible. This section investigates the thresholds in more detail, using the dynamic buckling pressure thresholds for $\delta = h/4$ as an example.

### (a) Centre-stable manifold of the saddle

Considering the spectrum of the linearization in the saddle at $\delta = h/4$ and $p/p_C = 0.512$ in figure 3, we see that the saddle has one stable eigenvalue and one unstable eigenvalue. Their respective eigenvectors correspond to directions in which trajectories exponentially converge to or diverge from the saddle. Without damping, the saddle appears as linearly neutral in all other directions: but with a little non-zero damping these other directions would be stable modes with trajectories converging to the saddle. This implies that close to the saddle the set of all initial conditions that do not diverge rapidly from the saddle forms a hypersurface, splitting the phase space near the saddle into two subsets (and the boundary hypersurface). One subset contains those initial conditions that buckle immediately. The other subset contains those initial conditions for which trajectories do not buckle immediately but instead oscillate around the node and either ultimately buckle or possibly converge to the nearby node if damping is present. The boundary is the set of all initial conditions whose difference to the saddle is spanned by the eigenvectors corresponding to the stable and neutral directions (directions that are neutral without damping become weakly stable with damping). Mathematically this boundary set is called the centre-stable manifold (CSM). Close to the saddle this CSM is approximately a hyperplane, a linear space of co-dimension 1, that is, of dimension one less than the entire space ($N-1$ in our simplified argument). Further away from the saddle the CSM is no longer a hyperplane but a (differentiable) curved $N-1$-dimensional hypersurface. It is known that CSMs of saddles can fold back on themselves dramatically, even in low-dimensional systems, allowing the system to become chaotic [22,23]. The CSM of the saddle depends on the parameter $p$ (as does the location of the saddle itself).

The sketch in figure 10 shows the geometry that we have been discussing in a heuristic three-dimensional projection from a notional $N$ dimensional phase space. The base plane shows the well-known 2D phase portrait of a one-degree-of-freedom system generated as a saddle and a node approach one another to give a saddle-node fold (or limit point). For easy visualization, the trajectory heading towards the node in this plane has been given a much higher damping level than we are currently discussing. The single axis normal to the base plane has to represent all the other phase dimensions. The first important sub-space of this third axis is the CSM which is illustrated as a green transparent upright surface (dimensionality, $N–1$), containing the saddle equilibrium and its intersection with the base plane (shown in yellow). This manifold acts as a threshold for buckling, as we can see by following the three adjacent trajectories coloured purple, red and blue all of which are heading towards the orange centre manifold of the saddle equilibrium point (a subset of the transparent green CSM and discussed further below). As shown by the vertical dashed lines, the blue trajectory lies behind the green manifold, and eventually diverges to the right implying the buckling of the shell. Meanwhile the purple trajectory lies in front of the green manifold, and ends up turning toward the unbuckled node equilibrium.

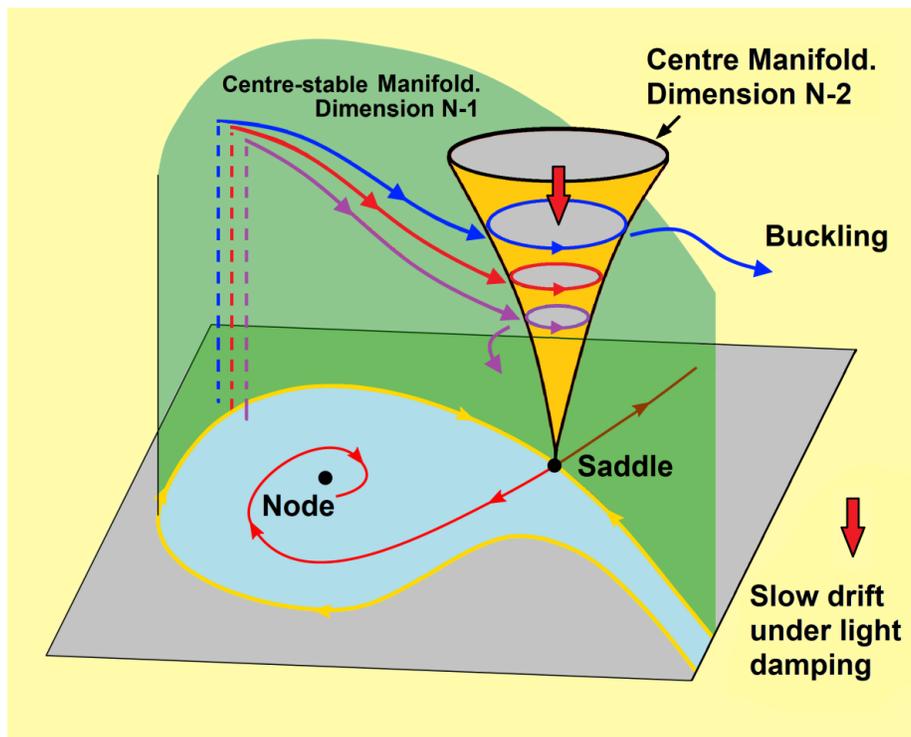

Figure 10: Geometric arrangement of thresholds and buckling trajectories in the full phase space (including position and velocities).

Somewhere between these two typical trajectories lies the special red trajectory which lies precisely in the green CSM. Initial conditions behind the surface generate immediate buckling, while those in front do not. However, the purple trajectory will make another round trip around the node. As the (green) CSM folds strongly further as it extends away from the saddle, after its next round trip the purple trajectory may lie behind those further folds of the green manifold, leading to multiple thresholds.

The final divergence of the trajectories is intimately related to the second important sub-space, the so-called centre manifold drawn in orange. This centre manifold is a subset of the green CSM, but has been expanded in the 3D projection of figure 10. When there is no damping this multi-dimensional manifold (dimension $N$–2) contains a variety of periodic and quasi-periodic orbits. On the introduction of damping the orbits inside this manifold drift slowly downwards towards the saddle equilibrium point. All trajectories close to the threshold (the green CSM) are caught up in these circling motions before they are thrown off in opposing directions, including the blue, red and purple examples in the sketch.

Since dynamic simulations create a small amount of damping, a simple criterion for the different sets is the long-time limit of the pole deflection $w_0(t)$. Let us denote by $w_{0,s}(p)$ the value of the pole deflection $w_0$ of the saddle equilibrium at pressure $p$ (transparent surface in left panel of figure 6). Then, (noting the pole deflections of interest are negative) a trajectory after pressure ramp to $p$

1. buckles, if $w_0(t) < w_{0,s}(p) - h$ for large times $t$.

2. avoids buckling, if $w_0(t) - w_{0,s}(p)$ converges to a positive value for large times $t$ (namely $w_{0,n}(p) - w_{0,s}(p)$, where $w_{0,n}(p)$ is the pole deflection of the unbuckled stable equilibrium at pressure $p$),

3. is on the threshold (in the CSM), if $w_0(t) - w_{0,s}(p)$ goes to zero for large times $t$.

With the small (and physically necessary) damping created by the dynamic simulations, the convergence to $w_{0,s}(p)$ will be slow for threshold trajectories (being slower for smaller damping). After an initial exponential approach to the CSM, damping will introduce a drift to the saddle equilibrium, which is the point of lowest energy in its own CSM. For zero damping, the CSM contains periodic and quasiperiodic orbits. These orbits are all themselves of saddle type (thus unstable), and, hence, not visible in dynamic simulations.

*(b) Thresholds as connections to the centre-stable set of the saddle*

Despite the slow convergence, the above distinction provides a simple criterion for establishing thresholds more accurately than shown in figure 6. A pressure ramp to low $p$ leads to a trajectory of the non-buckling type, while for ramps to $p = p_s$ the trajectory will buckle. Thus, for fixed end time $t_E$ we may apply a bisection in ramp pressures $p$ to find a pressure $p$ that leads to a trajectory that has $w_0(t_E) = w_{0,s}(p)$.

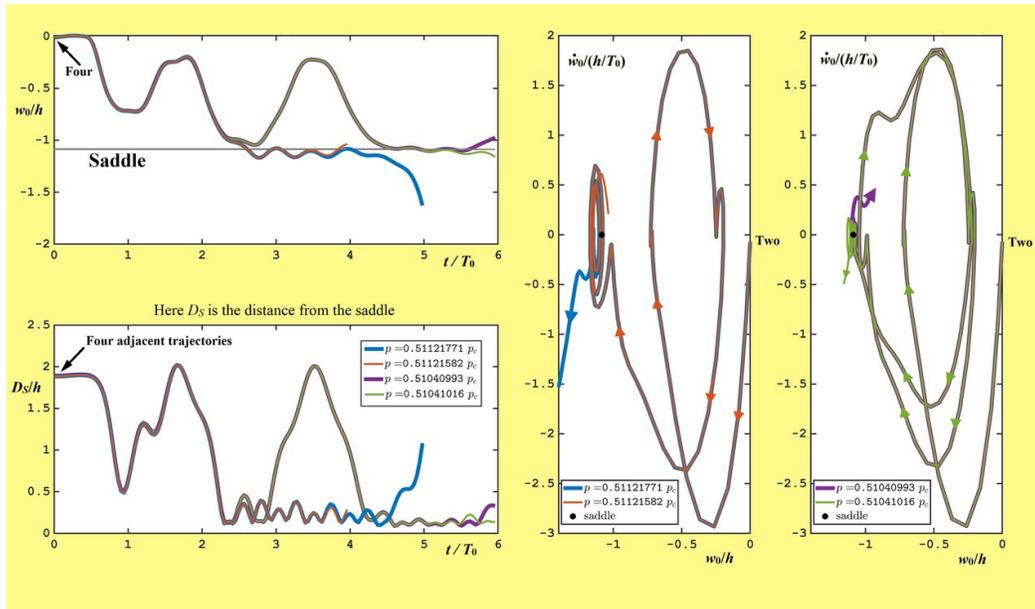

Figure 11. (Left, top) Time profiles $w_0(t)$ for 4 different pressure ramps, 2 near $p = 0.5104 p_c$ (green, purple) and 2 near $p = 0.5112 p_c$ (blue, red). (Left, bottom) distance of same trajectories from saddle as function of time. (Right) Same trajectories in $(w_0(t), \dot{w}_0(t))$ plane. Other parameters: $\delta/h = 0.25$, time stepsize $\Delta_t = 0.087 T_0$.

Figure 11 shows the bracketing trajectories for the result of the bisection for $t_E = 4T_0$ (blue and red), and $t_E = 6T_0$ (green and purple), for imperfection $\delta = h/4$ and time step $\Delta_t = 0.087 T_0$. As the top left panel of figure 11 shows, the pole deflection $w_0(t)$ performs oscillations around the saddle value $w_{0,s}(p)$ (grey horizontal line) for considerable time before $t_E$ (larger than $T_0$). During this time the trajectory is close to the saddle (as the bottom left panel shows). Hence, we can draw a first conclusion that the saddle computed for figures 1 and 3 indeed plays a key role in the buckling. However, the buckling threshold is not given by a trajectory that connects to the saddle, but rather a trajectory that connects into the CSM of the saddle (into a small amplitude periodic or quasi-periodic motion near the saddle). Panels on the right of figure 11 show the threshold-bracketing trajectories in their projection to the $(w_0(t), \dot{w}_0(t))$ plane. This projection also shows how the threshold trajectories make a small number of excursions where $w_0(t)$ is between $w_{0,s}(p)$ and 0 before reaching the CSM. The difference between the two bracketing pairs is the time it takes before reaching the CSM. The red/blue pair brackets the threshold trajectory reaching the CSM before $4T_0$ (during the initial time up to $2.5T_0$), while the green/purple pair reaches the CSM only after time $4.5T_0$. Since the threshold pressures used in figure 11 are close to each other, the threshold trajectory for $t_E = 6T_0$ is nearly identical to the threshold trajectory for $t_E = 4T_0$. All trajectories shown in figure 11 only diverge from each other while spending time near the saddle as small oscillations that are part of the CSM: see the near-periodic orbits in the projected phase portraits on the right panels of figure 11, and how previously nearly identical trajectories diverge from these small oscillations. The diverging trajectories follow opposite directions in the unstable dimension (the *outset* [22]) of the small amplitude oscillation in the centre manifold. This can be seen by comparing the end pieces of the red versus the blue (left $(w_0(t), \dot{w}_0(t))$ projection in figure 11), or green versus purple trajectories (right $(w_0(t), \dot{w}_0(t))$ projection).

*(c) Time ordering of buckling threshold trajectories and pressures*

From these observations we expect that there is a discrete sequence of buckling thresholds when considering a range of step load pressures $p$. The sequence is ordered by the time it takes for the threshold trajectory to get close to the CSM of the saddle. This order is not necessarily the same order as in the pressures $p$. For example, between the two thresholds pressures shown in figure 11 there may be more threshold pressures, for which the trajectory reaches the passive set much later in time (especially for small damping).

Thresholds that do not take a long time (such as the threshold given by the blue and purple trajectories in figure 11 for $t_g = 4T_0$ depend only moderately on the damping (that is, they have a well-defined limit for zero damping). However, the number of thresholds increases as the damping goes to zero, as additional thresholds may occur later and later in time. A rough estimate how many additional thresholds to expect for a particular damping level can be obtained by observing the amplitude and energy level of the small oscillations in the CSM that the first threshold trajectories converge. In figure 11 the small amplitude oscillations for the second (green/purple) pair near the saddle are already much smaller than the oscillations of the first (red/blue) pair. Thus, we expect at most one more threshold occurring after the two observed in figure 11 (in time ordering).

## 7. Conclusions

Accurate calculations for the buckling of elastic spherical shells under step pressure loading, based on small-strain/moderate rotation theory, have revealed nonlinear features of the dynamic buckling of imperfection-sensitive structures that seem not to have emerged in earlier studies. The most notable is the fact that there is not necessarily a clear threshold between pressure levels that cause buckling and those that do not result in buckling. Instead, particularly for a shell with a relatively small imperfection, there exists a cascade of buckling thresholds. The cascade of thresholds is sensitive to structural damping. For the spherical shell, and probably for other imperfection-sensitive shell structures as well, it appears that, the smaller the damping, the smaller the lowest pressure at which buckling occurs. For the spherical shells with the realistic damping levels employed in this paper, the lowest step buckling pressures were reduced by about 30% below the corresponding static buckling pressures for shells with relatively small imperfections (c.f., figure 8). For shells with larger imperfections, which buckle statically below about 60% of the classical buckling pressure of the perfect shell, the lowest step buckling pressures are reduced by less than 10% below the corresponding static buckling pressures.

These dynamic step buckling trends for the spherical shell differ significantly from trends predicted using simple 1-DOF imperfection-sensitive models. The simple models suggest than nearly perfect structures will buckle under step loads only slightly below the corresponding static buckling load, and that the ratio of the step buckling load to the corresponding static load *increases* as the imperfection increases. We have also found that the lower bound (the astatic pressure) on step buckling pressure for the spherical shell based on overcoming the energy barrier associated with the saddle of the energy landscape lies far below the computed step buckling pressure, especially for shells with small imperfections. By contrast, the lowest step buckling load of the simple 1-DOF models coincides with the astatic load. Since the damping in our dynamic simulations was non-zero and the computed lowest step buckling pressure of the spherical shells depends on damping, it remains an open question as to whether shells with no damping might ultimately after long periods of dynamic oscillation buckle at pressures just above the astatic pressure.

The oscillatory interaction between the so-called breathing mode and the modes contributing to buckling, first investigated for ring buckling in [18], appears to be ubiquitous. In some of the buckling literature this type of interaction is referred to as 'parametric resonance' [7]. For spherical shells buckling at the lowest step pressures, this interaction amplifies the modes contributing to buckling interactions to the point where snap buckling takes over. At step pressures sufficiently above the lowest buckling threshold, snap buckling can occur almost immediately without the preliminary oscillatory interactions.

Because the lowest buckling threshold depends strongly on damping, the damping in the simulations should be calibrated to the particular experimental situation studied. Our simulations suggest that most energy loss occurs at the breathing frequency such that matching the damping to observations at the breathing frequency is more important than the particular damping model. The lesson from the present study is that damping is an important consideration in the determination of the lowest step buckling threshold, because lowering the damping level adds additional thresholds that cause buckling with larger delay after the pressure step.

**Funding Statement** J. Sieber's research was supported by funding from the European Union's Horizon 2020 research and innovation programme under Grant Agreement number 643073, by the EPSRC Centre for Predictive Modelling in Healthcare (Grant Number EP/N014391/1) and by the EPSRC Fellowship EP/N023544/1.